\documentclass[11pt,twoside]{article}
\usepackage{asp2010}

\resetcounters

\begin{document}

\title{A PIONIER and incisive look at the interacting binary SS Lep}
\author{N.\ Blind$^1$, H.M.J.\ Boffin$^2$, J.-P.\ Berger$^2$,  J.-B. Lebouquin$^1$, and A. M\'erand$^2$ \\
\affil{$^1$UJF-Grenoble 1/CNRS-INSU, Institut de Plan\'etologie et d'Astrophysique de Grenoble (IPAG) UMR 5274, Grenoble, France \\
$^2$European Southern Observatory, Santiago, Chile}}

\begin{abstract}Symbiotic stars are eccellent laboratories to study a broad range of poorly understood physical processes, such as mass loss of red giants, accretion onto compact objects, and evolution of nova-like outbursts. As their evolution is strongly influenced by the mass transfer episodes, understanding the history of these systems requires foremost to determine which process is at play: Roche lobe overflow, stellar wind accretion, or some more complex mixture of both. We report here an interferometric study of the symbiotic system SS Leporis, performed with the unique PIONIER instrument. By determining the binary orbit and revisiting the parameters of the two stars, we show that the giant does not fill its Roche lobe, and that the mass transfer most likely occurs via the accretion of an important part of the giant's wind.
\end{abstract}

\section{Introduction} 
\label{part:intro}

SS Leporis (HD 41511) contains an evolved M6\, III giant and an oversized A1\, V star (see Tab.\ \ref{tab:param} for a list of its known characteristics), and presents the so-called Algol paradox, as the most evolved star is the least massive of the two, indicating mass reversal, through mass loss and mass accretion. The possibility that the system is indeed undergoing mass transfer is backed up by observations of regular outbursts \citep{struve_1930} as well as from the UV activity from the shell surrounding the A star \citep{polidan_1993}, while the presence of a circumbinary dusty disc \citep{jura_2001} hints at the fact that the process is not conservative. In this sense, the recent observations of \citet{verhoelst_2007}, indicating that the M star potentially fills its Roche lobe, is of great value.

We report here new interferometric observations in $H$ and $K$, obtained with the AMBER and PIONIER \citep{berger_2010} instruments attached to the Very Large Telescope Interferometer (VLTI).
%By nature, an interferometer samples the spatial frequencies of the source, so that with baseline length up to 130\,m, the VLTI gives access to spatial resolutions up to 1\,milli arcsecond (mas) in the near infrared (i.e.\ at least twenty times the resolution of an 8-m telescope). 
These observations allowed us to directly probe the most inner parts of SS Lep, and to unambiguously determine its morphology at different epochs, giving rise to a refreshing view of this system. In this contribution, we focus our study on the binary system, determining its orbit as well as the M star diameter. We will finally show that the giant does not fill its Roche lobe, in a strict sense.

\begin{table}[t!]
\begin{center}
 \begin{tabular}{lrcrcr}
\hline
                        			& System 				&& Previous 		&		& This work         \\
\hline
  $d$ [pc]        			& $279\pm24$ &[1] 		&		&				&			 \\
  $P$ [days]    			& 260.3 $\pm$ 1.8  &[2] 	&		&				& 			\\
  $f(m)$	     			& $0.261\pm0.005$ &[2] 	&		&				&			 \\
  $e$               			& 				&	&	 0.024 $\pm0.005$ &[2]	& 0.005 $\pm 0.003$			\\
  $i$           	     			& 				&	& $30^\circ \pm 10^\circ$  & [3] 	& $143.7^\circ \pm  0.5 ^\circ$\\
  $\theta_A$ [mas]   		&				&	& $0.53\pm0.02$  & [3] 		& 			\\
   $\theta_M$ [mas]   		& 				&	& $3.11\pm $0.32  &[3]			& 		2.296 $\pm$ 0.007	 \\
  $M_A$ [${\rm M}_\odot$] 	& 				&	& 2 $\sim$3 & [2] 				& 		2.71 $\pm$ 0.27		 \\
  $M_M$ [${\rm M}_\odot$] 	&				&	&  0.4$\sim$ 1 &[2] 			&  			1.30  $\pm$ 0.33 \\
  $1/q = M_A/M_M$            & 				&	& $4\pm1$ & [2] 				&	2.17 $\pm$ 0.35		 \\
\hline
 \end{tabular}
 \caption{\label{tab:param} Relevant parameters of the SS Lep system, as previously determined or from this work: $d$ is the distance, $P$ the orbital period, $e$ the eccentricity, $i$ the inclination, and $f(m)$ the mass function. For the stars, $\theta$ is the apparent diameter, and $M$ the mass. References: [1] \citealt{leeuwen_2007}; [2] \citealt{welty_1995}; [3] \citealt{verhoelst_2007}.
 % (4) \citealt{blondel_1993}; (5) \citealt{jura_2001}.
 }
\end{center}
\end{table}

%
%%%%%%%%%%%%%%%%%%%%%%%%%%%%%%%%%%%%%%%%%%%%%%%%%%%%%%%%%%%%%%%%%%%%%%%%%%%
\begin{figure}[b!]
	\centering
	\includegraphics[width=.33\textwidth]{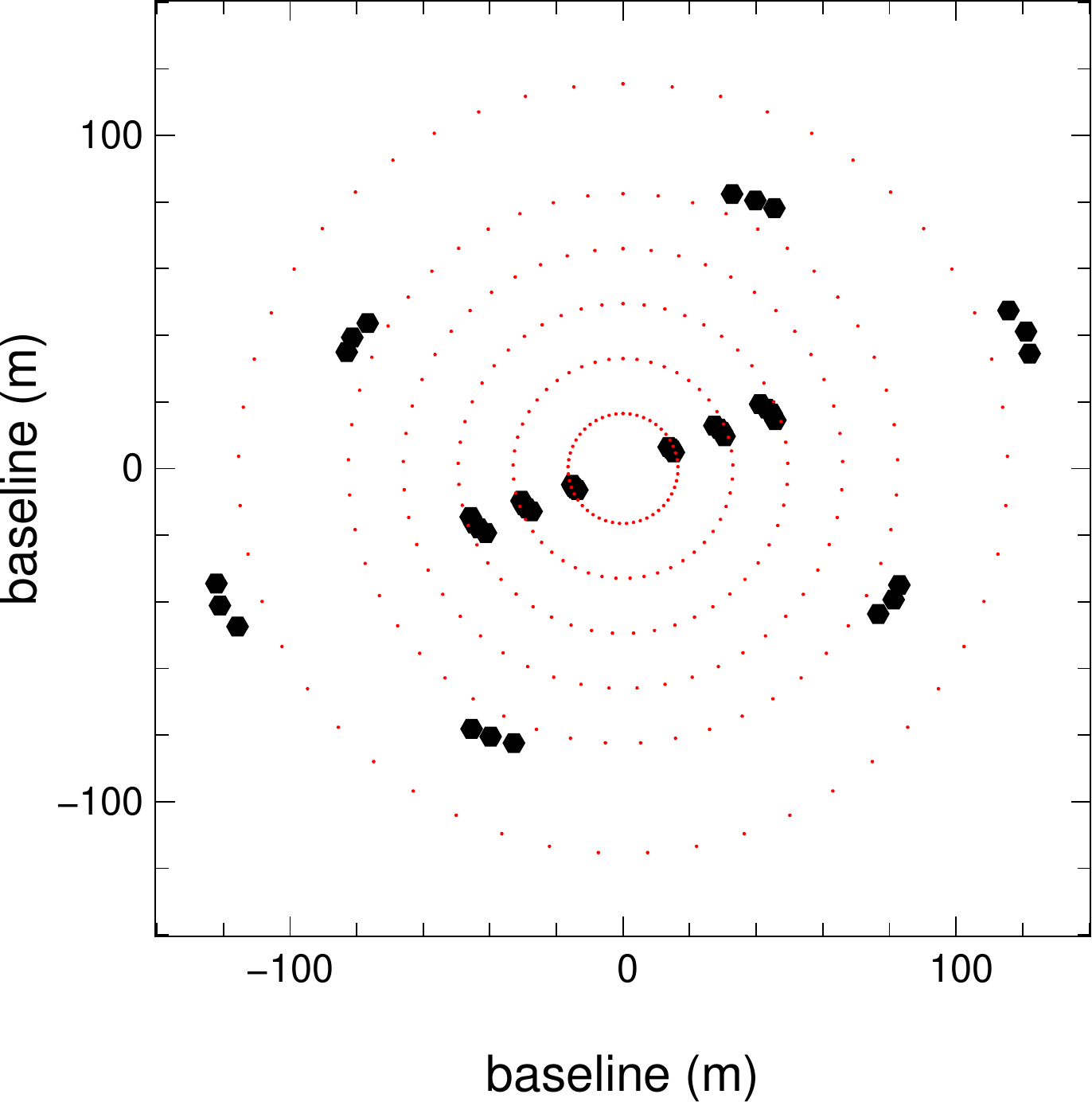}
	\includegraphics[width=.33\textwidth]{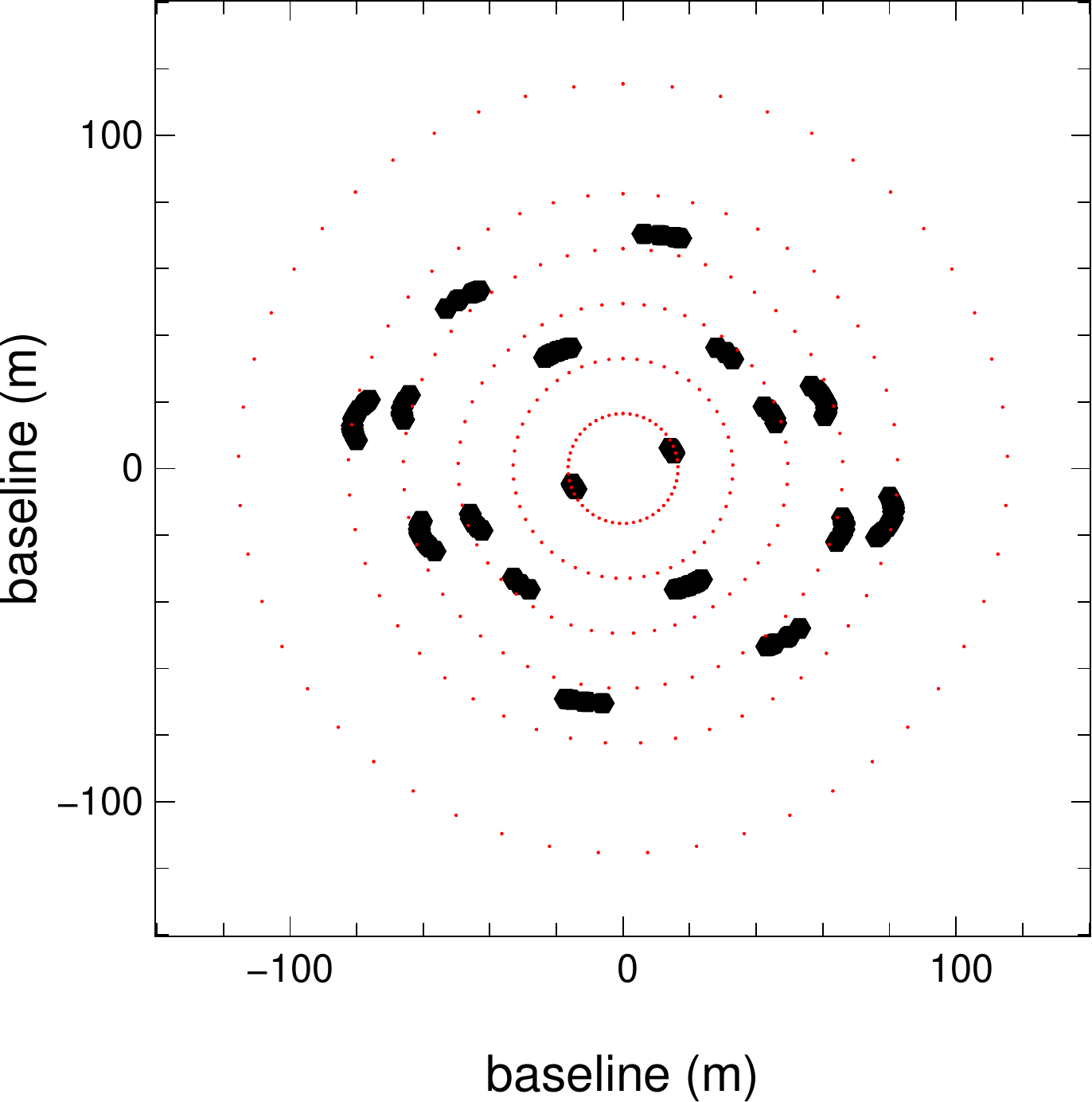}
	\caption{Typical  $(u,v)$-plane coverage of SS Lep as obtained with AMBER (left) and PIONIER (right). Each point represents a spatial frequency sampled by the instrument. The concentric dotted circles represent projected baselines lengths equal to 15, 33, 50, 65, 80 and 115 metres \label{fig:UV_plan}}
\end{figure}
%%%%%%%%%%%%%%%%%%%%%%%%%%%%%%%%%%%%%%%%%%%%%%%%%%%%%%%%%%%%%%%%%%%%%%%%%%%
%

\section{Observations, data reduction and modeling} \label{part:obs}

Data were collected at the VLTI with the three-telescope spectro-interferometer AMBER and the four-telescope visitor instrument PIONIER. Note that thanks to its additional telescope, PIONIER samples the spatial frequencies twice as fast as AMBER. AMBER data were obtained during 4 different nights in a period of 200\,days (more than half an orbital period) between November 2008 and April 2009. PIONIER data were obtained between October and December 2010 during the commissioning runs of the instrument. AMBER observations cover the $H$- and $K$-bands, while PIONIER only covers the $H$-band. For both, we have access to a low spectral resolution $R\sim40$ which multiplies the number of spatial frequencies sampled. This brings a wealth of information for the parametric modeling, especially for AMBER for which the $(u,v)$-plane coverages were relatively poor (i.e.\ the number of spatial frequencies sampled is low). Typical  $(u,v)$-plane coverages for AMBER and PIONIER can be seen in Fig.\ \ref{fig:UV_plan}.

The data clearly show that SS Lep is a spatially resolved binary whose M giant is also resolved in all observations and can be modeled as a uniform disc, whereas the circumbinary material is modeled as a gaussian envelope. We tried to detect a possible tidal distortion of the giant or matter escaping from its atmosphere by modeling it with an elongated uniform disc. Results were not conclusive as we lack the longest baselines required to measure distortions of the order of a few percents. The spatial resolution was not sufficient to resolve the putative shell or an accretion disc around the A star. Given the longest baselines of our observations (130\,m with a spatial resolution around 1\,mas), this agrees with the 0.5\,mas size estimated from the Spectral Energy Distribution (SED) in \citet{verhoelst_2007}. The model we used to fit the interferometric data therefore comprises six degrees of freedom: the relative flux contribution of two components of the system; the binary separation and its orientation; the size of the M giant and the size of the circumbinary envelope. The lack of short baseline information prevents us to constrain the spatially extended emission due to the large circumbinary disc ($FWHM\sim12$\,mas). We therefore focus on the binary in the following.

%
%%%%%%%%%%%%%%%%%%%%%%%%%%%%%%%%%%%%%%%%%%%%%%%%%%%%%%%%%%%%%%
\begin{figure}[t]
	\centering
	\includegraphics*[width=0.95\textwidth]{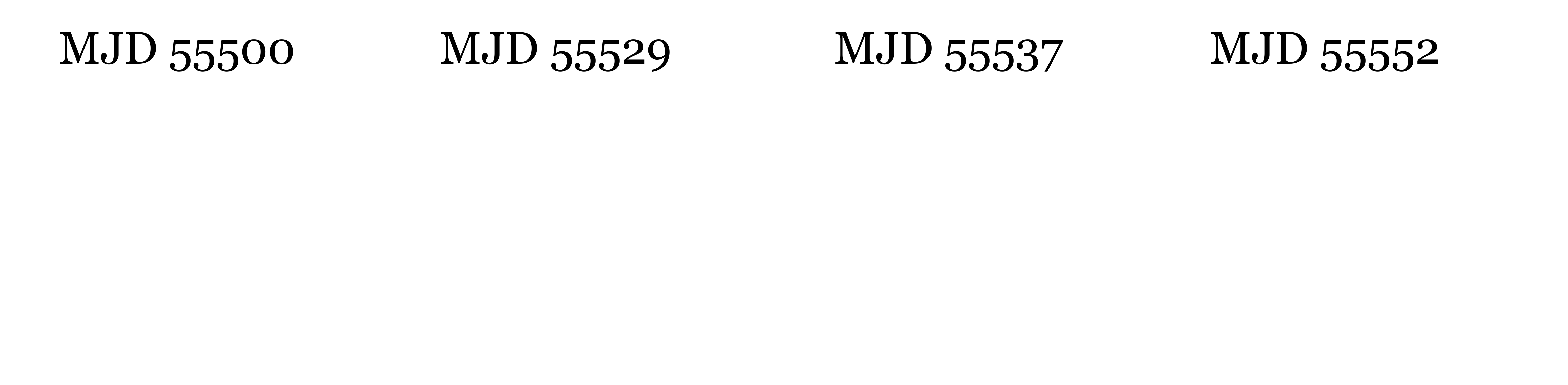}
	\caption{Model independent image reconstructions of SS Lep obtained during the four runs with PIONIER. The resolved M giant (about 2 mas large) and the A star are clearly identified, and separated by roughly 5\,mas. The distortion of the giant in the images is most certainly due to an asymmetric PSF rather than to a real tidal effect. A few faint artefacts are visible on the periphery of the images. \label{fig:SSLep_images}}
\end{figure}
%%%%%%%%%%%%%%%%%%%%%%%%%%%%%%%%%%%%%%%%%%%%%%%%%%%%%%%%%%%%%%
%

\section{Image synthesis} \label{part:image}

With its four telescopes, PIONIER allowed us to perform a reliable model independent image reconstruction of SS Lep with the MIRA software \citep{thiebaut_2008} for the four observations (see Fig. \ref{fig:SSLep_images}). Each image clearly shows the binary nature of SS Lep, the separation being slightly smaller than 5\,mas. From one observation to the other, we can observe the rotation of the system. 

The A star and its shell have an expected spatial extension of 0.5\,mas \citep{verhoelst_2007} so that we do not expect to resolve them with the VLTI baselines. Therefore, the size of the spot corresponding to the A star defines more or less the point spread function (PSF) of the image, about 1\,mas. The M giant being the most luminous component of the system in the $H$-band, we identify it on images as the darkest spot. With respect to the A star, we clearly see that it is spatially resolved and measures approximately 2\,mas in diameter, while the orbit size is roughly 5\,mas. We expect the distortion observed on the images to be due to an asymmetric filling of the $(u,v)$-plane (implying a non-circular PSF on the reconstructed image) rather than to a real tidal distortion. As a matter of fact, the tidal distortion would be around 7\%, i.e.\ less than seen in the images. Additionally, its orientation in the images corresponds well with the asymmetry observed in the corresponding $(u,v)$-planes. It was actually not possible to image the circumbinary disc because of the lack of data with short baselines.

\section{The binary} \label{part:binary}

\subsection{The orbit of SS Lep}

\begin{figure}[t]
	\centering
	\includegraphics[width=.5\textwidth]{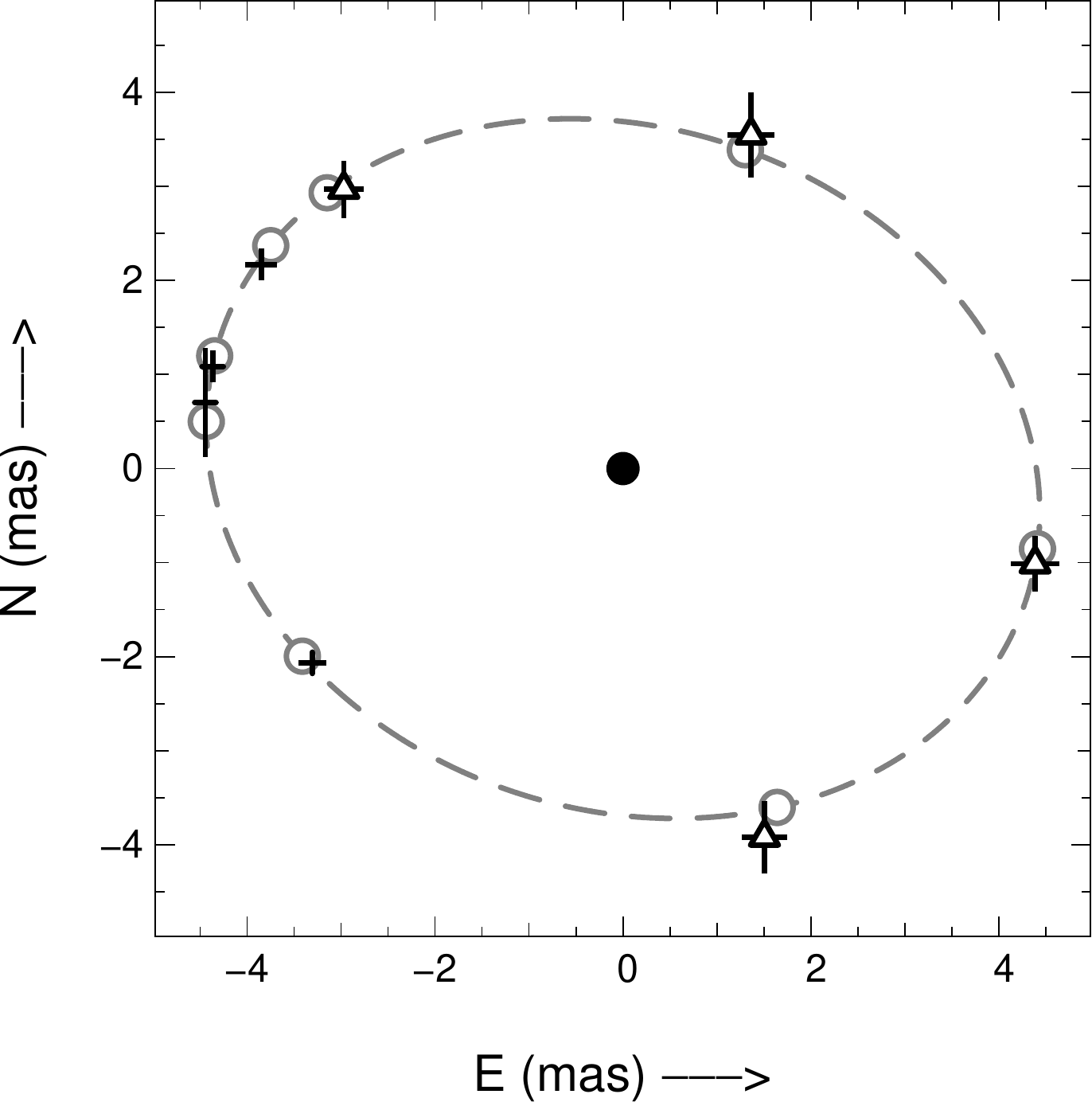}
	\caption{SS Lep's best orbit (dashed line) obtained by combining previous radial velocities \citep{welty_1995} with our astrometric measurements. The A star is indicated as the dot in the centre. AMBER and PIONIER points are represented by the triangles and crosses, respectively, with the error bars. The corresponding points on the best orbit are the closest circles from each point. 
	\label{fig:orbit}}
\end{figure}

To compute the most reliable orbit possible, we combined the eight astrometric positions of the binary (separation and orientation of the stars) obtained from our interferometric observations, with the radial velocities of \citet{welty_1995}. The best-fit orbit is shown in Fig. \ref{fig:orbit}, and the derived orbital parameters are listed in Tab. \ref{tab:orbit}. The errors on the orbital elements are estimated via Monte-Carlo simulations. The inclination angle of $143.7^\circ$ and the close to zero eccentricity are in agreement, but more precise, than the previous estimates by  \citeauthor{welty_1995}. This result definitely invalidates the periastron-passage mass transfer scenario of \citet{cowley_1967}, which required an important eccentricity ($e=0.134$) to explain the regular outbursts of the system. 
%
%%%%%%%%%%%%%%%%%%%%%%%%%%%%%%%%%%%%%%%%%%%
%%%%%%%%%%%%%%%%%%%%%%%%%%%%%%%%%%%%%%%%%%%
%
%
%%%%%%%%%%%%%%%%%%%%%%%%%%%%%%%%%%%%%%%%%%%%%%%%%%%%%%%%%%%%%%
\begin{table}[hb]
\begin{centering}
	\begin{tabular}{ccccc}
	\hline
 Semi major & Inclination & Eccentricity & Longitude of the                  & Argument of \\
   axis $a$  & angle $i$   & $e$            & ascending node $\Omega$ & periapsis $\omega$ \\
	\hline
$4.492\pm0.014$\,mas & $143.7\pm0.5^\circ$ & $0.005\pm0.003$ & $162.2\pm0.7^\circ$ & $118\pm30^\circ$  \\
	\hline
	\end{tabular}
	\caption{Orbital parameters of SS Lep obtained by combining previous radial velocities \citep{welty_1995} with our 8 astrometric measurements.
	\label{tab:orbit}}
\end{centering}
\end{table}
%%%%%%%%%%%%%%%%%%%%%%%%%%%%%%%%%%%%%%%%%%%%%%%%%%%%%%%%%%%%%%
%

\subsection{The distance and the masses}

Combining the orbital parameters with the binary mass function estimated by \citet{welty_1995}, we can estimate the individual mass of the stars, and, thus, the mass ratio. The main source of uncertainty in this estimate resides in the distance, as determined by Hipparcos. Using the distance and the angular separation of the two stars, we obtain $a = 1.26 \pm0.06$\,AU, and thus, through Kepler's third law, the total mass of the system is estimated as $4.01\pm 0.60  \,{\rm M}_\odot$. Thanks to the inclination that we determined and to the binary mass function, we can derive the individual masses and the mass ratio: $M_A = 2.71\pm0.27\,{\rm M}_\odot$, $M_M=1.30\pm0.33\,{\rm M}_\odot$, and $1/q = M_A/M_M = 2.17\pm0.35$.

\subsection{The M star diameter}

We measure a uniform disc diameter for the M star $\theta_{M, UD} = 2.208\pm0.012$\,mas. The previous VINCI observations of \citeauthor{verhoelst_2007} led to a higher value of $2.94\pm0.3$\,mas, most likely because theirs was the result of a one-year survey of the source, without any phase information in the interferometric data. This involved to model the system as a symmetric object, and to mix up the interferometric signatures of the rotating binary with the resolved giant one. Depending on the authors (\citealt{hanbury_1974}, \citealt{davis_2000}, or \citealt{verhoelst_2007}), the conversion factor to a limb-darkened diameter differs by a few percents and we fix it to 1.04. This leads to a limb-darkened diameter equal to $\theta_{M, LD} = 2.296\pm0.013$\,mas.
Our results is still in very good agreement with the limb-darkened diameter estimated from the SED of \citeauthor{verhoelst_2007} ($\theta_{M,LD} = 2.66\pm0.33$\,mas). 
Taking into account the uncertainty on the distance, the M giant radius is thus $R_M = 69 \pm 3\,{\rm R}_\odot$ (almost twice as small as previously thought), and the surface gravity is therefore $\log g \sim 0.8$.

%%%%%%%%%%%%%%%%%%%%%%%%%%%%%%%%%%%%%%%%%%%
\begin{figure}[t]
	\centering
	\includegraphics[width=.4\textwidth]{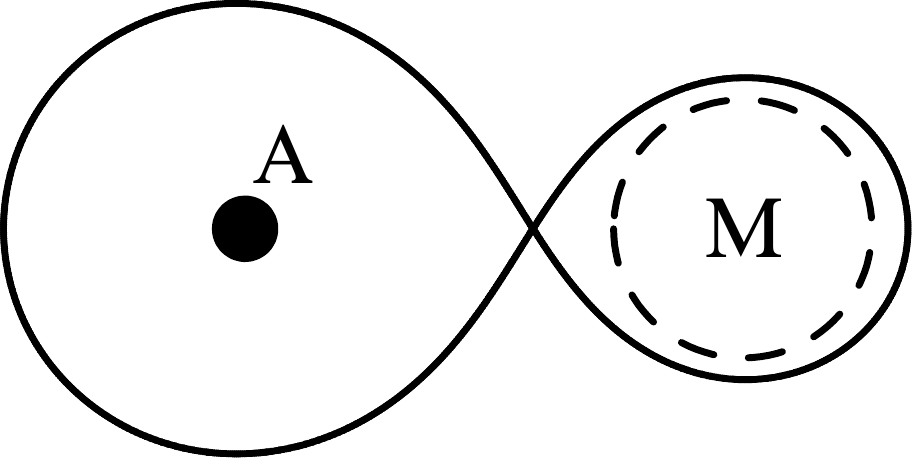}
	\caption{Representation of the Roche equipotential (solid line) for a mass ratio $1/q = 2.2$. The limb-darkened diameter of the M giant is in dashed line, and the A star one is the dark dot.  \label{fig:RocheEquipotential}}
\end{figure}
%%%%%%%%%%%%%%%%%%%%%%%%%%%%%%%%%%%%%%%%%%%

\section{The mass transfer process} \label{part:masstransfer}

The circumbinary disc and the envelope around the A star, as well as the activity seen in the ultraviolet, are good hints for the  occurrence of mass transfer. The disc additionally indicates that the process is not conservative. With the parameters derived above, we are able to show that, conversely to previous results, the M giant only fills about $85\pm3\%$ of its Roche lobe (Fig. \ref{fig:RocheEquipotential}), and the mass transfer is more likely due to the accretion of the M giant wind. If we assume the star's surface to follow an equipotential surface, we expect a tidal distortion of the giant from 5 to 7\%. The wind speed being lower than the orbital one ($v_{\rm orb}=48$\,kms$^{-1}$) the system is possibly in the particular case of a wind Roche lobe overflow \citep{mohamed_2007}, where a substantial part of the stellar wind can be accreted. From the simulations of \citet{nagae_2004} the wind accretion efficiency for these velocities is around 10\% for a mass ratio equal to 1. The accretor in SS Lep being the most massive component, the efficiency could be significantly higher. However we do not expect that all the wind can be accreted because the circumbinary disc is a proof of the non conservative behaviour of the mass transfer process. Assuming the M giant looses $10^{-7}\,{\rm M}_\odot$yr$^{-1}$ \citep{dupree_1986} through wind, the accretion rate onto the A star should be between $10^{-8}$ and $10^{-7}\,{\rm M}_\odot$yr$^{-1}$. It is thinkable, however, that some enhanced mass loss takes place in such kind of systems, and this is possibly only a lower limit. This is still much lower than the value of $10^{-4}\,{\rm M}_\odot$yr$^{-1}$ quoted by \citet{verhoelst_2007}.

\section{Conclusion and future work}

We have presented here the results of our observations, and we focused on the binary. After having computed the characteristics of the orbit, we demonstrated that the mass ratio is lower than previously thought and that the M giant does not fill its Roche lobe. However the system is in a configuration where a substantial part of the giant's stellar wind can be accreted by the A star. The refreshing and precise view of SS Leporis given by these observations paves the way for the study of interacting binaries with interferometry, in combination with spectroscopic studies.

The interferometric data also gave us information about the relative luminosity of the three components of the system (the M and A stars, and the circumbinary disc). There is still some work to fully exploit these data. In particular, we lack a low resolution spectrum of SS Lep between $1.6$ and $2.5$\,$\mu$m to compute the absolute luminosity of each component and extract more specific information. The current data also present good hints of unmodeled material escaping the system. We plan to observe SS Lep with the full high spatial resolution capacity of the VLT, with NACO and PIONIER to study in more details the binary and the circumbinary disc, and extract as much information as possible about the mass transfer and the interactions. We also plan to use AMBER to observe the M giant wind and detect an asymmetry due to the high efficiency of its accretion onto the A star.

\acknowledgements{It is a pleasure to thank O.\ Absil and G. Dubus for their help.}

%%%%%%%%%%%%%%%%%%%%%%%%%%%%%%%%%%%%%%%%%%%%%%%%%%%%%%
%%                 REFERENCES                       %%
%%%%%%%%%%%%%%%%%%%%%%%%%%%%%%%%%%%%%%%%%%%%%%%%%%%%%%

\bibliographystyle{asp2010}

\end{document}